# The Story about One Island and Four Cities.

# The Socio-Economic Soft Matter Model - Based Report.

**Agata Angelika Rzoska[1*] and Aleksandra Drozd-Rzoska[2*]**

[1.] University of Economics, Dept. of Marketing, 1 Maja 50 St., 40-257 Katowice, Poland

[2] Institute of High Pressure Physics Polish Academy of Sciences,

ul. Sokołowska 29/37, 01-142-Warsaw, Poland

* Correspondence: A.A. Rzoska; agata.rzoska@edu.uekat.pl; A. Drozd-Rzoska: arzoska@unipress.waw.pl

**Abstract:** The report discusses the emergence of the Socio-Economic Soft Matter (SE-SM) as the result of interactions between physics and economy. First, demographic changes since the Industrial Revolutions onset are tested using Soft Matter science tools. Notable is the support of innovative derivative-sensitive and distortions-sensitive analytic tools. It revealed the Weibull-type powered exponential increase, with notably lesser rising rate since the crossover detected near the year 1970. Subsequently, demographic (SE-SM) patterns are tested for Rapa Nui (Easter) Island model case and for four large 'hallmark cities' where the rise and decay phases have occurred. They are Detroit and Cleveland in the USA, Łódź (former textile industry center), and Bytom They (former coal mining centers) in Poland. The analysis explicitly revealed scaling patterns for demographic changes, influenced by the historical and socio-economic backgrounds, and the long-lasting determinism in population changes. Universalistic features of demographic changes are discussed within the Socio-Economic Soft Matter concept.

1. Introduction

Nicolai Copernicus, Galileo Galilei, Isaak Newton, … created the Modern Scientific Method. It is related to the non-speculative approach, with the experimental cross-verification of heuristic and thought concepts and the verbalization of results using a mathematical apparatus. It isn't easy to distinguish one of these giants, but just Isaac Newton finally shaped the Modern Science path (Westfall, 1994). It led to unprecedented progress associated with new fundamental knowledge and innovations. The world entered a qualitatively new era of innovation-driven development, removing many problems that have plagued people for millennia. It is the Industrial Revolutions times (Groumpos, 2021).

Isaac Newton is particularly known for two monographs defining the basics of Mechanics (dynamics), Gravity, and Optics. He introduced the concepts of derivatives, differential equations, and integral analysis for a coherent and in-depth description. A particular impact had the indication that seemingly separate phenomena may have a common 'universal' source. Who before expected that the falling apple and the motion of planets and comets can describe the same universal law of gravity, concluded in a simple, functional equation? (Westfall, 1994)

Adam Smith, the 'father' of modern economics, was so overwhelmed by the concept of 'all-embracing universality' that he began to look for such phenomena in social and political economy-related phenomena (Smith, 1776 & 2009). It is worth noting Newton was also directly involved in such problems. Recognizing the great merits of his mind, Newton was nominated the Royal Mint head. His main task was preventing (dramatic) currency value falls. Newton quickly realized that this was mainly due to a (very) physical action - scraping the edges of the coins, which reduced their weight. Newton solved the problem by developing a new hard metal alloy for coins and a serrated edge for coins. Thanks to a multidisciplinary innovation combining physics, materials engineering, economics, and behavioral observations, Newton solved the grand socio-economic problem of the Kingdom. This innovation is still in use. (Westfall, 1994)

This reports attempts to apply some basic physics, particularly soft matter physics, issues to describe socio-economic and demographic peculiarities, as well as possible universalistic concepts behind them.
To validate the application of physics to such issues, let's look at Newton's famous laws of motion from a socio-economic point of view:

*1.* When no force is acting on the body, or when these forces balance each other, it is at rest or moves at a constant speed in a straight line. *Does it not also mean proportional, 'linear' development of the company, city, and even the state when positive and negative external factors balance each other?*

*2.* The motion is related to the force $F$ acting on a body with mass $m$, the measure of its inertia of inertia: $F = am$, where $a = const$ is the acceleration. *In socio-economics, an example may be the difficulty of changing the activities of a large company with enormous structural inertia compared to the rapid pro-market returns of medium and small companies?*

*3.* There is no single force. Any action is accompanied by a reaction. *The number of analogies in economics and everyday life is evident in the given case.*

Three decades ago, Pierre Gilles de Gennes was awarded the Nobel Prize in Physics (1991) for another grand unification. He introduced the new Soft Matter (SM) category linking many materials common in our surroundings. At first, it included polymers, liquid crystals, micellar systems, critical liquid,… (de Gennes & Badoz 1996). Now, within Soft Matter also, bio-systems, including colonies of bacteria and viruses, or food as the very complex soft matter case are considered (Drozd-Rzoska et al., 2002; Rzoska et al. 2011; Drozd-Rzoska et al. 2005a; Drozd-Rzoska et al., 2005b; Drozd-Rzoska, 2005; Drozd-Rzoska 2006; Drozd-Rzoska, 2009). For the latter, one can recall popular products from milk, yogurt, mayonnaise, ketchup…. to pastes, cakes, and meat (Mezzenga et al.). The Soft Matter concept has also been extended to topological (Serra et al., 2020) and quantum (Thedford et al., 2022) systems, offering new insight into fundamental Space properties.

What does Soft Matter mean? (de Gennes & Badoz, 1996). What can connect such apparently distinct systems? Only two common features are essential for universal characterizations and scaling patterns:

A. the dominance of collective, mesoscale 'structures', mainly correlated assemblies of atoms, molecules, or any other type of entities building a system/material

B. extreme sensitivity to exo- and endogenic impacts, which in fact, is the consequence of point (*A*).

In 2007, *The Guardian* magazine farewell Pierre Gilles de Gennes in following words (Goodby, Gray, 2007):"…*Newton of our time has died aged 74. He was awarded Nobel Prize, the Lorentz medal and Wolf prize for "discovering that methods developed for studying order phenomena in simple systems can be generalized to more complex forms of matter*.'

In the recent work, the authors of this report put forward the thesis that one can consider the Socio-Economic Soft Matter, due to the exemplary fulfillment of the above (*A, B*) general conditions. The global population $P(t)$ was chosen as a model system, and its evolution from the beginning of the Holocene / Anthropocene era, with the onset at $t_{ref.} = 12000 BC$ was examined using Soft Matter tools (Rzoska, 2016; Rzoska & Drozd-Rzoska, 2023). The following scaling pattern for the population changes was revealed:

$$P(t) = p_0 exp\left[\pm \frac{\Delta t}{\tau(t)}^{\beta}\right] \quad (1)$$

where $P(t)$ denotes population changes as the function of time, $p_o$ is for the prefactor, $\tau_i$ is the processing time constant, $\beta$ is the power exponent; $\Delta t = t - t_{ref.}$, and $t_{ref.}$ is for the reference (onset) time; $\tau$ denotes the relaxation time; '$\pm$' is related to the rise and decay, respectively.'

In Soft Matter systems, the exponent $\beta = 1$ indicates that the system is governed by a single relaxation time concerning dynamic processes or modes. For $0 < \beta < 1$, Eq. (1) is named as the stretched relaxation relation, indicating a distribution of relaxation time (processes). In physicochemical systems $\tau(t) = \tau = const$. Values related to $\beta > 1$ are for the 'compressed' relaxation processes, which hardly occur in physico-chemical systems. The exception is $\beta = 2$, which is equivalent to the normal (Gaussian) distribution [2] of probability for different random allowed states of the system. Notable that the parallel of Eq. (1) resembled one of Weibull's functions are used for general probability distribution analysis for dynamic processes via different relaxation channels. It is often implemented in various technological applications, as well as in medicine and microbiology.

For the simplest case, $\beta = 1$, data analysis based on Eq. (1) can be concluded using the following linear dependence:

$$lnP(\Delta t) = lnp_0 - \frac{1}{\tau}\Delta t \quad (2)$$

The single-relaxation time dynamics ($\beta = 1$) validates the linear behavior for the plot

$y = lnP(\Delta t)$ vs. $x = \Delta t$ or simply $x = t$, with the slope related to the relaxation time reciprocal. The relaxation time $\tau$ denotes the time required to the change of $P(t)$ by $1/e \approx 0.3678$. ...factor. The convenient metric can be $\tau_{1/2} = \tau \times ln2$ describing the time required for 50% change in $P(t)$. For the complete 'powered' Eq. 1 with the exponent $\beta \neq 1$ one should consider $(1/e)^{1/\beta}$ decay (Rzoska & Drozd-Rzoska, 2023). The nonlinear fitting is used for describing 'experimental' data by Eq. (1) with the powered exponential function related to the exponent $\beta \neq 1$. Practical implementations in soft matter systems showed that the direct nonlinear fitting using Eq. (2) leads to surprisingly large errors for derived parameters. This basic problem can be avoided when considering the derivative of Eq. (1) Rzoska & Drozd-Rzoska, 2023):

$$\frac{dlnP(t)}{dt} = \frac{dP(t)/P(t)}{dt} = \pm \frac{\beta}{\tau} \Delta t^{\beta-1} \qquad (3)$$

and subsequently, the transformation of data based on the following equation (Rzoska & Drozd-Rzoska, 2023):

$$y(t) = log_{10}\left|\frac{dlnP(t)}{dt}\right| = log_{10}\left(\frac{\beta}{\tau}\right) + (\beta-1)log_{10}\Delta t =$$
$$= \left[log_{10}\left(\frac{\beta}{\tau}\right) - log_{10}\tau\right] + (\beta-1)log_{10}\Delta t = A + B \times x \qquad (4)$$

The plot $y(t) = log_{10}[dlnP(t)/dt]$ vs. $x = log_{10}\Delta t$ should yield the linear dependence with the slope $s = \beta - 1$, i.e. the exponent $\beta = s + 1$, in the domain where Eq. (1) can be applied.

Using Eq. (4), one can reveal regions governed by different values of the exponent $\beta$ without the nonlinear fitting via Eq. (1). It also shows dominant local trends and distortions. Notably that the analysis via Eq. (1) supported by the preliminary analysis based on Eq. (4) reduces the final fir via Eq. (1) to only one parameter, in a strictly defined allowed data domain.

Recalling the issue of the human population evolution, which is the crucial topic of the given report, the most significant model was introduced by Malthus, who also declared the inspiration from Isaak Newton legacy. He assumed the constant rate of population changes and applied the derivative analytic tool introduced by Newton (Malthus, 1798; Weil & Wilde, 2010; Kaack & Katul, 2013):

$$:\frac{\Delta P(t)}{\Delta t} \to (\Delta t \to 0) \to \frac{dP(t)}{dt} = rt \quad \Rightarrow \quad P(t) = p_o exp(rt) \qquad (5)$$

where $r = r(t) = const$ is the Malthus rate coefficient.

Eq. (5) correlates with 'powered' Eq. (1) for $\beta = 1$ and $t_{ref.} = 0$.

Malthus confronted the exponential population rise suggested by Eq. (5) with an expected linear increase of available food resources at that time. The inevitable intersection of the trends of changes in the population and the number of resources indicated the unavoidable lack of the latter and, consequently, the certainty of famine and social disorder.

In the mid-19[th] century, Velhulst supplemented the Malthus model focusing on the population growth in restricted resource conditions leading to the rise of the population and its subsequent stabilization. Such a situation can be functionalized as follows (Velhulst, 1847&2022):

$$\frac{dP(t)}{dt} = rt\left(1 + \frac{P(t)}{K}\right) \Rightarrow P(t) = \frac{K}{1+exp(-rt)(K-p_o)/p_0} = K\left[1 + exp(-rt)\frac{K-p_0}{p_0}\right]^{-1} \qquad (6)$$

where *K* characterizes the number of available resources

The third type of the Malthus-type models is related to the function Eq. (1):

$$P(t) = p_0 exp(rt^\beta) \qquad (7)$$

It is applied via non-linear fit, generally in an arbitrarily selected time domain, which is always associated with large errors of derived parameters. The Weibull model function is often cited as a heuristic justification.

The generalized Weibull-type Eq. (1) (Rinne, 2008), supported by the distortions-sensitive and derivative-based analysis defined by Eq. (4) was applied to get new insight into the global population growth. It covered the extreme period from the Holocene epoch onset.

**Figure 1** presents the resume of these results, focused on the period after 1800, which covers the interest of the given report. The result presented in Fig. 1 shows the non-monotonous pattern of changes based on Eq. (1), governed by changes in the value of the exponent $\beta$.

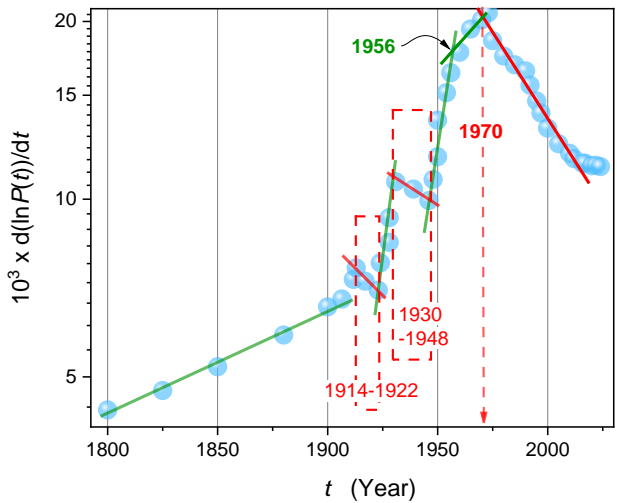

**Figure 1** The plot showing the evolution of the World population using the Weibull-type Eq. (1) and the distortions-sensitive analysis of its applicability (indicated by linear domain) via Eq. (4). Slopes of lines are coupled to the power exponent $\beta$ in the Weibull-type Eq. (1) $slope = s = \beta - 1$. The behavior described by $s > 0$ (*green lines*) is for the 'strong' rising rate of the population. The behavior described by $s < 0$ (*red lines*) is for a significant decrease in the rising rate of the population. The crossover $s > 0 \rightarrow s < 0$ indicates the inflection between mentioned trends. Characteristic years related to such changes are indicated. Based in ref. (Rzoska & Drozd-Rzoska, 2023).

Notable are periods where the increase of the global population slowed down associated with World Wars: WWI and WWII. Notable that for the latter such trend started near 1930, which can be linked to the Grand Economic Depression and terminated at 1948. The global population rise pattern definitely changed from a fast to a slow rise of the global population near the year 1970. Is it the impact of 'cultural revolts' in 1968 and its consequences? The Global Population has the feature that is often expected for physical models – isolation from disturbing external factors, or at least the knowledge and control about them.

This work implements the concept of socio-economic Soft Matter for selected, characteristic, smaller-scale human populations, where the 'isolation' seems to be difficult, if possible. Population changes in a size-restricted territory reflect internal factors, such as political and social conditions and available 'attractive' resources from food to energy. The latter can also be non-material, as shown below. The basic discussion of this report starts from the Rapa-Nui (Easter Island case), which can be considered the canonic model for developing a 'small', isolated human population.

2. **Population Changes: Microbiology parallel and the Rapa Nui (Easter Island) case**

In the first centuries of the 1$^{st}$ Millennium, one of human history's most significant exploration adventures occurred. Polynesians started grand oceanic travels, leading to the settlement of islands in the vast expanse of the Pacific Ocean. The success of these efforts has remained unprecedented. It occurred on small but nautically perfect boats - catamarans and trimarans ensuring excellent journey stability. The exploration and settlement expeditions set off to other islands thousands of kilometers away without knowing whether they existed at journey terminals. The enormous knowledge acquired at that time also supported the success of Captain James Cook's great expedition on the Endeavor ship, guided by Polynesian navigator Tupaya. The farthest island reached by explorers was Rapa Nui island, which European explorer Admiral Jacob Roggeeven called Easter Island, in honor of his arrival there on Easter Sunday, April 5$^{th}$, 1722. At that time, it was inhabited by 2 - 3 thousand people. Later studies showed that in the 16$^{th}$ and 17$^{th}$ centuries, the population of Rapa Nui was as high as 10-15,000 inhabitants. The next visit of Europeans was also of a research nature: in 1770, the expedition of Captain Felipe Gonzales de Ahedo, on the order of the viceroy of Peru, explored the island for 5 days. In 1774, Captain James Cook visited the island, during the grand Endeavor ship expedition. The visit resulted in excellent maps, descriptions of nature, comments regarding inhabitants, and communication on 'monumental' Moai statues on the coast. Captain Cook estimated the population between 700 and 200 inhabitants, with the latter being more likely. Between 1722 and 1770 on Rapa Nui, the final stage of the conflict between two existing clans known as 'short-ears' and 'long-ears took place. It led to the disappearance of the 'long-ears' clan, so completely that no reliable genetic material remained. The remaining verbal tradition of the story on Rapa Nui indicates that the 'long-ears' may have been the ruling clan of the Island (O'Leary, 2021).

A unique feature of the development of the human population on Rapa Nui, associated with the formation of an extraordinary civilization with extraordinary achievements, was operating in conditions of isolation from the outside world, both from other Pacific islands and mainland South America.

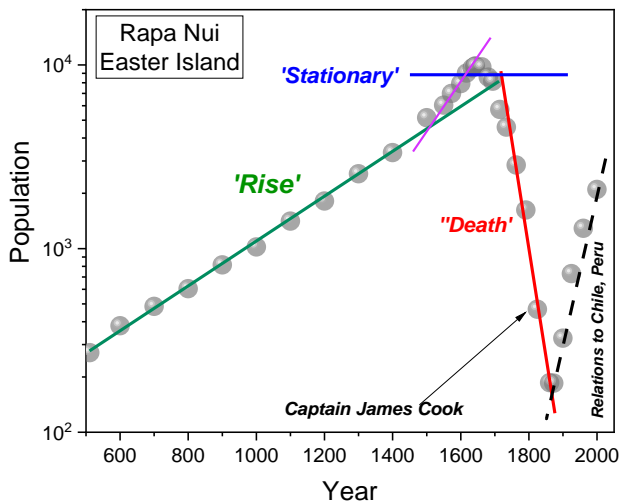

**Figure 2** Changes of Easter Island (Rapa Nui) population since the hypothetical settlement till nowadays, based on estimations from ref. (O'Leary, 2021' Stevenson et al., 2006).

The settlement of Rapa Nui is associated with the last phase of the great oceanic expansion of the Polynesians. Shortly afterward, climatic conditions deteriorated periodically, which additionally affected the limitation of expeditions, which took place on small and relatively fragile boats. In a small area of Rapa Nui, there was also rapid deforestation, especially the disappearance of palm trees. It meant no boat building for ocean voyages or even fishing surrounding Rapa Nui was possible. The inhabitants were cut off from the world's richest fisheries, and food resources. They only had resources related to the interior of the island.

Recently, there have been indications that residents have developed 'ecological' gardens inside the island, significantly solving the resource problem. Rapa Nui's population has grown to about 10,000. However, from the mid of the 16$^{th}$ century, the global cooling of the climate, referred to as the Little Ice Age, took place. In Europe, its impressive manifestation was the winter freezing of the Baltic Sea, and ice travels from Poland to Sweden. On Rapa Nui, it is a time of enormous intensification of the construction of increasingly huge and monumental Moai sculptures - forged in quarries inside the island and transported on wood logs to the coast. Such activities must have absorbed the entire relatively small population and led to the final deforestation. Suddenly a violence-related collapse led to a rapid depopulation, the disappearance of one of the clans, and the destruction of the existing civilization. It can be speculated that the intensification of activities related to the construction of Moai was an attempt at a mystical response to the deteriorating climatic conditions. These efforts probably resulted in rebellion, internecine wars, and finally, the collapse of structures, civilization, and the population's disappearance. **Figure 2** presents population changes of Rapa Nui population in the semi-log scale, directly recalling the exponential portrayal via Malthus Eq. (5) or Weibull-type Eq. (1) with the exponent $\beta = 1$. It is worth noting that there was an extra increase in population from about 1500 to the beginning of the 17th century. In Europe, it was a time of significant improvement in climatic conditions, which probably also had a global scale.

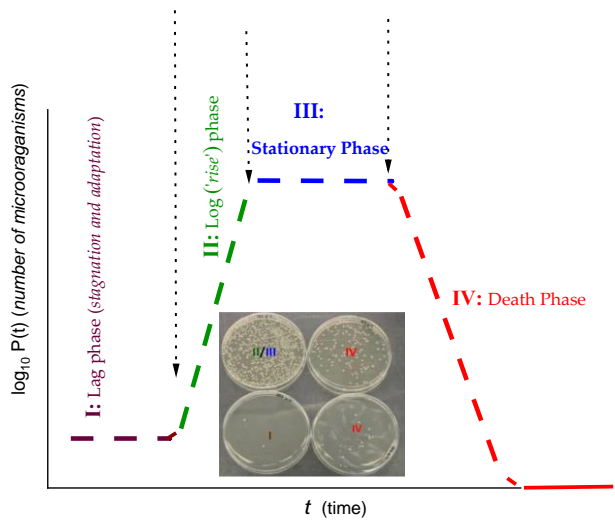

**Figure 3** The illustration of the evolution of microorganisms population development in a closed container with a defined (limited) amount of food. The semi-log scale facilitates the manifestation of the single-relaxation (Eq. (5) or Eq. (1) with the exponent $\beta = 1$) behavior, expressed via the sloped linear behavior. The photo presents the amount of E-Coli bacteria colonies in subsequent development stages (Rzoska et al., 2017).

The middle of the 17th century is already the fullness of the Little Ice Age, and strong cooling. On Rapa Nui there is a very short population stabilization period and, subsequently, a very rapid decline, which can be related to the comments presented above. However, this picture may be an oversimplification. The last decade studies have shown that Rapa Nui natives have developed advanced eco-friendly gardens in the central parts of the island. Data emerged that a giant tsunami wave also could sweep Rapa Nui. The importance of the above factors can be significant and can complicate the picture of Rapa Nui population changes.

**Figure 3** shows changes in the bacteria population in a container with limited food. The semi-log-scale reveals the Malthus-type rise (phase II) followed by the stationary phase III. Phases II and III together can be related to the Velhulst bimodal model (Eq. (7)). Finally, the population disappears (phase IV). The evolution of the bacterial population, which can also be considered the Active Soft Matter, is shown in Fig. 2. It is similar to the population change in Rapa Nui. However, there are some important differences. First is the extremely short stationary phase in Fig. 1. Second is the rise and decay asymmetry. It can be attributed to the fact that the development and survival of the human population on Rapa Nui were strongly related to internal self-ordering. Its destruction occurred in a 'rapid' manner and in a short period.

## 3. Population Evolution: hallmark post-industrial cities case

The question arises whether population changes evoking analogies to the dynamics of collective phenomena within Soft Matter Science may appear for human

'clusters' situations, where apparent interaction with the surrounding is obligatory. To discuss this issue, the population development of a few hallmark post-industrial cities in Poland and the USA is discussed, from the beginning of the 19th Century until now. Worth emphasizing is the differences between these countries. In the 19th Century, Poland did not exist, and its territory was divided between the three partitioning empires, Russia, Prussia / Germany, and Austria. For each of them, social and economic development followed a different path. Poland regained independence in 1918. It was one of the main arenas of World War I (WWI), leading to enormous destruction and impoverishment. Twenty years later, in 1939, World War II (WWII) broke out, causing Poland one of the greatest human and material losses. After the war, there was a forced change of the state borders and a forced link to the Soviet-Russian Empire. It terminated only in 1990.

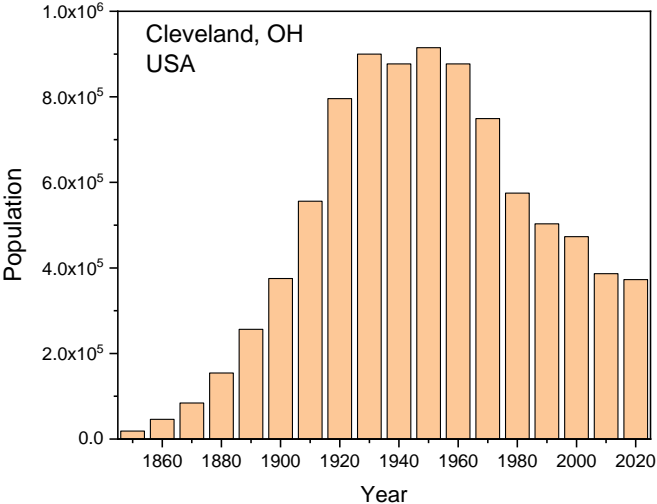

**Figure 4**  Population changes in Cleveland (OH, USA) in the classic bar presentation

In turn, the USA is a country which, from its beginnings throughout the 19th century, until now, is growing harmoniously. United States of America participated in WWI and WWII as the most important contributor. However, these wars occurred outside the territory of the USA, and surprisingly, the country was becoming more powerful and more prosperous, which is perhaps an unprecedented 'paradox' in history. In the USA, a country on a continental scale, excellent pro-development conditions have always existed, mainly pro-economic and pro-innovative, ... but also for personal freedom.

What does the development of selected, specific cities look like in the context of such dramatically different countries?
The main initial motive for their selection was the presence of a phase of strong population growth and then, after the stabilization phase, a rapid and significant decline in the population, which shows some similarity to patterns presented in Figs. 2 and 3.

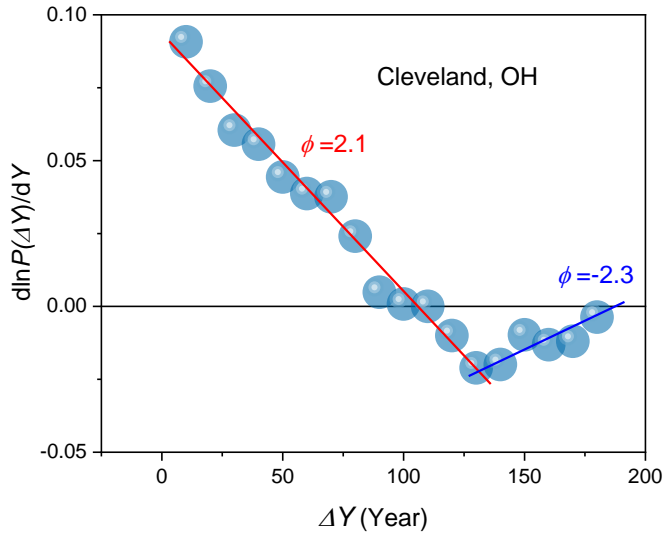

**Figure 5**  The linearized derivative-based analysis (Eq.(4)) for population data Cleveland OH, given in Fig. 4. Linear domains validate the portrayal via the powered exponential Eq. (3), with exponents given in the plot.

**Figure 4** shows population changes in Cleveland, OH (Census, US: Cleveland, 2022), in the semi-log scale recalling Malthus behavior. Cleveland is a large city and harbor on the great Erie Lake coast. It is an important transport hub, and this fact – associated with the location – has been significant for development since the onset. A significant driving force behind the development was also a variety of industries, primarily related to metallurgy. Cleveland has played the role of a considerable financial, commercial, and scientific center. The latter is exemplified by three universities and the space research center. One cannot forget the famous Cleveland Orchestra. Despite such exciting characteristics, the population changes presented via the standard bar representation (**Figure 4**) show the stationary period after the substantial increase and a significant population decrease.

**Figure 5** shows the distortions-sensitive analysis of population data from Fig. 4 via the linearized derivative Eq. (4). Linear domains validate the portrayal via the powered exponential Eq. (1), with exponents given in the plot. **Figure 6** shows the portrayal of population changes in Cleveland based on data presented in Fig. 4 and 5. The long 'rise' and 'decay' domains are notable, with the stationary period between years ~ 1930 – 1950. The development of Cleveland was significantly motivated by the activity as the hub and exchange center for goods, information, and finances. The main factor here was initially water transport on the huge reservoirs of the great lakes, stretching from the east to the far west of the USA. Rail transport rather complemented this, but car/lorry transport and the development of the motorways/highways network changed the situation. The convenient port location ceased to be a crucial pro-development factor.

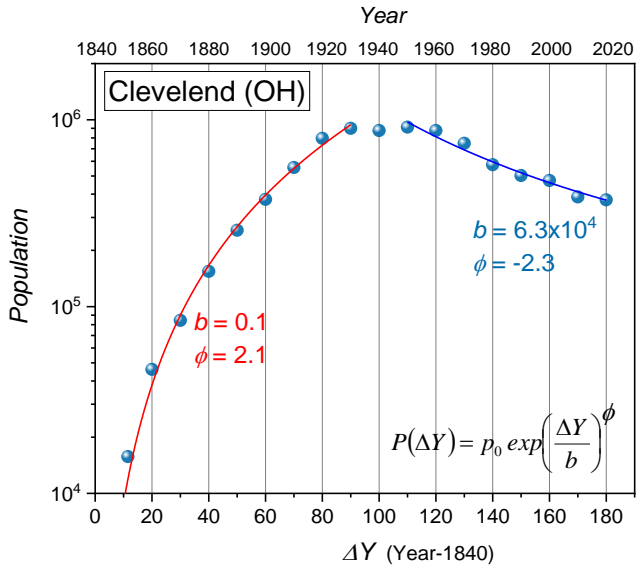

**Figure 6** The semi-log presentation of population data Cleveland OH (USA) with the powered exponential portrayal via Eq. (1). Note the link to Fig. 4.

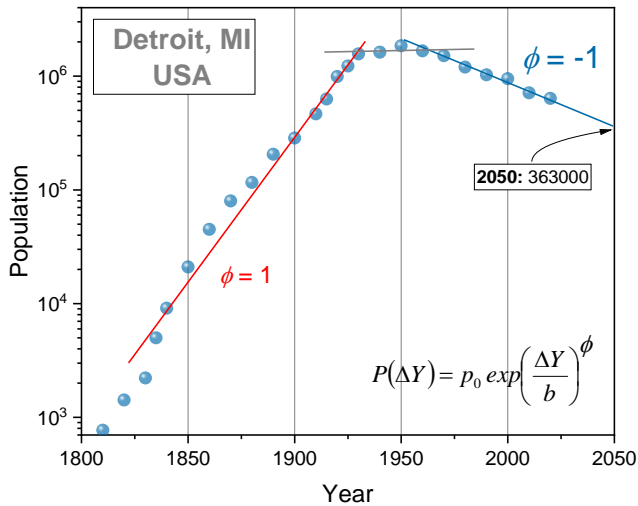

**Figure 7** The evolution of the population in Detroit MI (USA) in the semi-log scale. Solid lines are for the Malthus-type portrayal with the exponent $\beta = 1$ in Eq. (1).

**Figure 7** shows available population data for Detroit, MI (USA) (Census US, Detroit, 2022) in the semi-log scale. It reveals the excellent Malthus-type portrayal associated with the single-mode relaxation, linked to the exponent $\beta = 1$ in Eq. (1) particularly since 1900, which can be associated with the onset of the grand automotive industry in Detroit. Notable that a similar Malthus-type pattern associated with the exponent $\beta = 1$ in Eq. (1) also appears very distant from the case discussed above: Bytom in Poland.

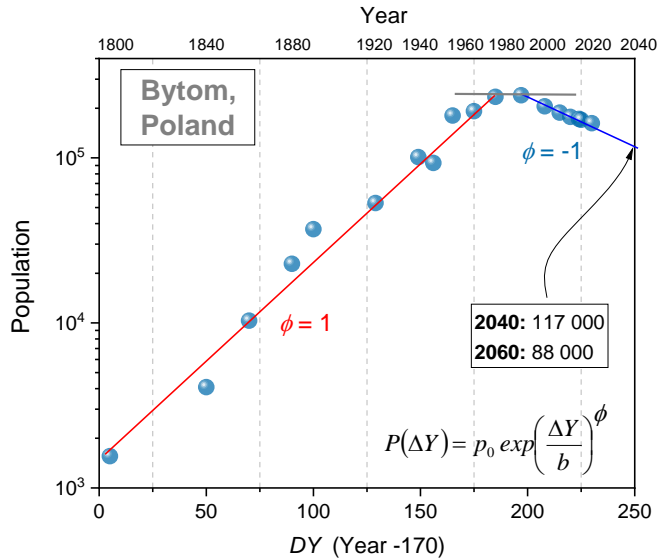

**Figure 8**  The evolution of the population in Bytom (Silesia, Poland) in the semi-log scale. Solid lines are for the Malthus-type portrayal with the exponent $\beta = 1$ in Eq. (1).

It is shown in **Figure 8**, based on data available in (Bytomski.pl, 2022). The Malthus-type population increase extends from 1800 until 1980. This trend exists despite (i) two world wars, (ii) changes in nationality: until 1918 Prussia / Germany and then Poland, (iii) the population exchange in 1945/1945 from German to Polish, (iv) the shift in the political & economic system from quasi-communist the real world of 'capital' economy. Bytom is an old city, but its rapid rise was associated with rich coal deposits. The reduction of the population can be associated with the rapid decline in the role of coal as a strategic energy resource. The decline is also Malthusian-type.

In Poland, one of the large symbolic cities with a declining population and various social and economic problems is Łódź, located approximately 140 km east of Warsaw. The history of this city is peculiar. It developed from a small village near the border between Russia and Prussia / Germany Imperia border. The encouragement introduced by the Tsar of Russia led to the creation of an exceptionally rapidly expanding textile center there, compared to Manchester UK, in the second half of the 19th century. **Figure 9** shows the extraordinary feature of the evolution: the population of Łódź increases linearly till the eighties, and subsequently, the explicitly linear decreases occur. It might suggest that the evolution is beyond the Malthus-type or powered-exponential patterns, which seemed to be universal (based on data avail at (Łódź w Liczbach, 2022)). Notable, till the year 1915 Łódź belonged to the Russian Tsar Empire, and the enormous economic capacity of the Empire stimulated its development. In the period 1905-1914, increased financial freedom boosted Russia's economy, which seems reflected in the extra population growth. In 1915 the extra population rise was associated with refugees, followed by the massive compulsory evacuation with the Russian army leaving the territory of Poland. The population increase before WWII can be linked to the enormous poverty after the Grand Crisis.

The population collapse during WWII (till 1946) can be associated with the holocaust of the Jews, a significant part of Łódź population, and refugees' outflow.

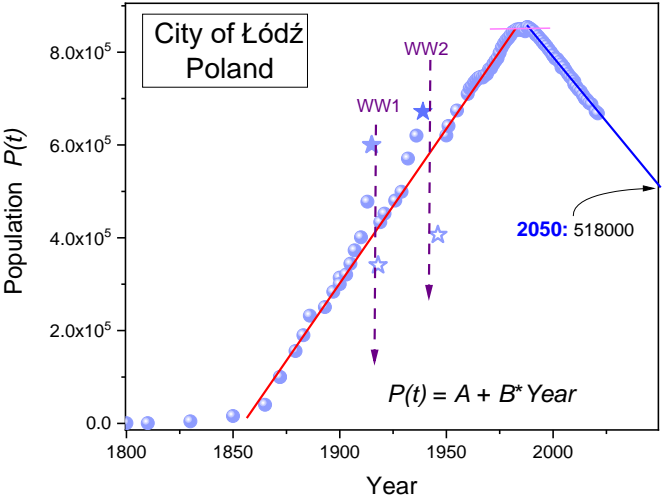

**Figure 9** The evolution of the population in Łódź (Poland). Note the explicit linear changes for the rise and decay of the population.

After the Second World War, Łódź returned to mass textile production for the benefit of the Russian Empire, this time in the form of the Soviet Empire, which ruled Poland through 'Polish communists'. This trend changed in the second half of the 1980s when the communist system collapsed. Since then, a permanent decline in the population - also linear and still lasting – has taken place

## 4. Conclusions

This work develops the socio-economic soft matter concept for discussing population changes in selected post-industrial hallmark cities. First, the reference case of Rapa Nui population, geographically isolated from any impact for centuries, was discussed. The analogy of the pattern observed for a colony of bacteria in a closed container with a limited amount of food/resources was indicated. For the Rapa Nui population development, the impact of global climate change and internal social ordering seems to be also significant. Subsequently discussed, post-industrial (nowadays) cities can be related to interactions with the environment. Nevertheless, the characteristics of their development tracked by population changes show noticeable similarities to the model case of Rapa Nui.

For Detroit, the uninterrupted and continuous "Malthusian" exponential growth was described by Eq. (1) with an exponent $\beta = 1$ took place. It suggests one dominant process/motivator of changes in dynamics, described by the single relaxation time. Regarding Detroit, the definition of this factor is obvious - it is the automotive industry. It also applies to the decline in the Detroit population that began as early as the mid-1960s as shown above. The decline in the population may result from a decrease in the demand for labor in the automotive industry and a relative reduction in salaries. The automation and the associated increase in productivity and increasing competition from Asian and European producers influenced the dominant industry in Detroit, and consequently, the population of this largely mono-cultural city has decreased.

Population changes in Cleveland (OH) are specific: both the rise and decay have the Weibull-type dependence described by Eq. (1) with the exponent $\beta \approx 2$  The city emerged as a significant transport and exchange hub due to its favorable location on the great Lake Erie coast. Its location, supported by various industries, has shaped the city's success for decades and led to strong population growth. However, since 1950, the population of Cleveland has continued to decline, which must also be reflected in economic and economic aspects. The 'powered' exponential Eq. (1) describes both the increase and the decrease, which suggests that it is a multi-channel process related to a set of relaxation times. Let us recall that the economy of Cleveland was multi-faceted from the beginning but dependent on a single factor - a great harbor. The weakening of this driving-force factor had to influence the development trends of other 'development channels', influencing the development and population. Since the mid-1950s, the final dominance of trucks and modern railways transport has taken place in USA. Notable that Detroit and Cleveland are cities (i) located in the USA, which is a country of model freedom and business support, (ii) it is characterized by unprecedented mobility of the workforce, (iii) the permanent increase of the US population has taken place.

The perfectly one-channel 'Malthusian' population dynamics is also emerging in Bytom (Poland), a city so distant from Detroit. In Bytom has occurred essential changes related to the nationality of citizens, state belonging, and the political system. They seem to be not significant for the population changes! One may conclude that only the driving force meaning huge deposits of high-quality coal, is important. The depletion of deposits and unfavorable long-term price changes in the international coal market from the mid-1980s limited the role of coal mining. The decay of the dominant economic stimulants started, resulting in a single exponential population decay in Bytom. It is worth recalling that it reflects the city's socio-economic attractiveness.

In Poland, Łódź symbolizes a city with some problems associated with political system changes. Like Detroit and Bytom, Łódź was the city created and shaped for decades by one industry related to large weaving and textile factories. Until the First World War and after the Second World War, the production of this industry was consumed, motivated by the enormous needs of the Russia Empire, does not matter governed by Tsar or 'communists'. At the end of the 1980s, this factor practically disappeared, and competitive and cheaper textiles from Asia appeared, which won the market in Poland. Soon later, weaving factories collapsed. Thousand of people lost jobs and the life-path concept. Often, they worked in textile factories for generations. New workforce positions emerged very slowly. It caused a huge and probably still existing trauma. One can expect that it was strengthened because factories employed mainly women.

As mentioned above, the city's population changes in a city can be associated with its economic attractiveness. The last factor should be treated not only in terms of salaries but also expenses, related to the life quality, price of housing, accessibility, and cost of social facilities, transport, … According to the author, the specific dynamics of population changes in Łódź can be explained by developing Eq. (1) in Taylor series, for $\beta = 1$:

$$P(t) = p_0 exp\left[\pm \frac{\Delta t}{\tau(t)}^{\beta=1}\right] \Rightarrow P(t) = p_0 \left[1 \pm \frac{1}{\tau}\Delta t \pm \frac{1}{2\tau}\Delta t^2 \pm \frac{1}{3\tau}\Delta t^3 \pm \cdots\right] \quad (8)$$

Thus, with relatively small values of the argument, or the influence of a parameter shaping a given trend, we get:

$$P(t) \approx p_0 \left[1 \pm \frac{1}{\tau}\Delta t\right] \quad (9)$$

that is the linear evolution of the population. Such a situation may occur when the impact of the dominant economic force on the surrounding is realized via non-interacting 'channels', with negligible feedback interactions between them. As a result, additional power terms in Eq. (8) are not activated, and the development of the city, also measured by the evolution of the population, is linear as in Eq. (9). In the opinion of the authors, the hypothetical minimal feedback effects that can create pro-development added value in Łódź can be associated with the unique situation that the dominated driving force were women, for all decades accepting worse working conditions, pay, and social environment. This situation is dramatically opposite to Bytom or the Upper Silesia region in general, where the authorities have always been afraid of the wrath of miners, not only men, but organized and decisive men. So, isn't the still difficult situation of Łódź a legacy of decades or even centuries of economic exploitation of women?

In summary, it can be stated that some aspects of the development of selected urban centers can be considered as a Socio-Economic Soft Matter entity, with the population dynamics described and explained under Soft Matter Science.. The environment seems to be a kind of averaged factor - in Soft Matter it can be described as 'mean field approximation', which always results in a qualitative simplification of the description of important processes. In such an approach, population changes are 'managed' not only by its internal example of bacteria in a container or Rapa Nui island but also by the existence of an 'attractor' in a given city, attracting people from outside the city, which in turn may start to leave the city when the attractor weakens, it disappears completely, or a much stronger other attractor appears nearby. In Poland, an example of such a new strong attractor can be Warsaw, located relatively close to Łódź, currently well connected, with attractive jobs, for various professions.

**Author Details:**
Agata Angelika Rzoska; ORCID: 0000-0003-4121-3152
e-mail: agata.rzoska@edu.uekat.pl
Aleksandra Drozd-Rzoska: ORCID: 0000-0001-8510-2388
e-mail: arzoska@unipress.waw.pl
**Competing Interests:** the Authors declare no competing interests

**Contribution:** the authors declare equal contributions regarding research and manuscript preparation issues.
**Data availability:** the authors declare the availability of data on requests to the authors.
**Funding**: This research is not related to any specific grant from funding agencies in the public, commercial, or not-for-profit sectors.